\documentclass[aip,jcp,preprint]{revtex4-1}
\usepackage{amsmath}
\usepackage{amsfonts}
\usepackage{amssymb}
\usepackage[pdftex]{graphicx}

\begin{document}

\title{Development of a New Parameter Optimization Scheme for a Reactive Force Field (ReaxFF) Based on a Machine Learning Approach}

\author{Hiroya Nakata}
\email{hiroya.nakata.gt@kyocera.jp}
\affiliation{R and D Center Kagoshima, Kyocera corporation, 1-4 Kokubu Yamashita-cho, Kirishima-shi, Kagoshima, 899-4312, Japan}
\affiliation{the authors equally contribute in this study}
\author{Shandan Bai}
\affiliation{R and D Center Kagoshima, Kyocera corporation, 1-4 Kokubu Yamashita-cho, Kirishima-shi, Kagoshima, 899-4312, Japan}
\affiliation{the authors equally contribute in this study}

\begin{abstract}
   Reactive molecular dynamics (MD) simulation is performed using a reactive force field (ReaxFF). 
   To this end, we developed a new method to optimize the ReaxFF parameters based on a machine learning approach. 
   This approach combines the $k$-nearest neighbor and random forest regressor algorithm to 
   efficiently locate several possible ReaxFF parameter sets, thereby the optimized ReaxFF parameter 
   can predict physical properties even in a high-temperature condition within a small effort of parameter refinement.
   As a pilot test of the developed approach, the optimized ReaxFF parameter set was applied to perform chemical vapor deposition (CVD) of 
   an $\alpha$-Al$_2$O$_3$ crystal.
   The crystal structure of  $\alpha$-Al$_2$O$_3$ was reasonably reproduced even at a
   relatively high temperature (2000 K). The reactive MD simulation suggests that the (11$\overline{2}$0) 
   surface grows faster than the (0001) surface, indicating that the developed parameter optimization technique 
   could be used for understanding the chemical reaction in the CVD process.
\end{abstract}


\maketitle



\section{Introduction}
   A chemical reaction that typically involves bond cleavage and formation
   plays an important role in the field of material science.
   In principle, it is possible to handle the chemical reaction by a 
   quantum mechanical (QM) approach, but the simulation of the chemical reaction 
   is still computationally challenging.
   Because expensive Hessian calculations\cite{hess1,hesspar} or 
   molecular dynamics (MD) simulations are needed for simulation of the chemical reaction,
   the QM applications are limited to within only 100 or 200 atoms 
   in the system, and a large-scale (more than 10,000 atoms) 
   reaction analysis is virtually impossible within the framework 
   of the standard QM approach.

   Two approaches are currently possible 
   for a large-scale reaction analysis based on the QM method: 
   a hybrid quantum mechanical/molecular mechanical (QM/MM) approach\cite{QMMM1,QMMM2,ONIOM} 
   and a fragment-based approach.\cite{frgrev,MCF,DC0,XPol,DC2,DC3,eemb2,MFCC3,PMISP,ADMA,EFP2013,pdf1}
   Although the majority of MD studies of large-scale reaction analysis employ the QM/MM approach, the boundary conditions between QM and MM are sometimes a problem. 
   Thus, MD based on QM/MM is not straightforward, and a challenging study of adaptive QM/MM molecular dynamics was recently reported.\cite{QMMMadaptive,watanabe2016adaptive} 
   In contrast, in the fragment-based approach, vibration analysis has been developed\cite{IMCMO,GEBFHESS,SMFHESS,MTAhess,QMQM,FMO_HESS}
   and a 10,000-atom enzyme reaction has become possible.\cite{FMOFD2}

   However, despite the above success of the QM-based approach, 
   MD simulations with QM are 
   limited to within only 1$\sim$100 ps.\cite{FMOMD}
   Thus, it is not yet practical to apply QM-based MD to 
   a chemical reaction in a material in which the system size is typically of nanometer scale 
   and nanosecond simulation time.

   One possible way to investigate the chemical reaction in nanoscale systems
   is currently the classical force field approach.\cite{van2001reaxff,EVBforce} 
   Because the classical force field approach has difficulty in handling the chemical reaction, the reactive force field (ReaxFF) approach was developed by van Duin et al.\cite{van2001reaxff}
   An additional advantage of ReaxFF is the ability to study a wide range 
   of material simulations, including solid-state crystal, molecular crystal, and gas molecules.
   Thus, ReaxFF has been applied to many chemical reactions in 
   materials.\cite{reaxff01,reaxff02,reaxff03,reaxff04,reaxff05}
  
   To apply the ReaxFF MD simulation of a chemical reaction, 
   force field parameter refinement is sometimes necessary.
   The force field parameter fitting is usually performed 
   to reproduce the static properties such as energy, force, and charge by
   using a particular QM training data set.
   To fit the ReaxFF parameters to the QM training data set, a
   genetic lgorithm,\cite{reaxFFGA01,reaxFFGA02,reaxFFGA03,reaxFFGA04,reaxFFGA05}
   and a multiobjective genetic algorithm\cite{GARFF,ReaxFFGAM}
   have been intensively studied; thus, force field parameter optimization based on the QM method has become relatively easy.

   The success of the ReaxFF simulation and force field fitting has opened up new possibilities for application in computational material science. However, it is not yet straightforward to apply ReaxFF to a new chemical reaction, in particular in a nonequilibrium process such as chemical vapor deposition (CVD). This is because, in a nonequilibrium process, there is uncertainty in the quantification of simulated quantities of interest,
   and the preparation of a QM training data set 
   is sometimes impossible for our quantities of interest. 
   Mishra {\textit {et al.}} improved the uncertainty quantification by including the dynamic approach to simulating reactive MD.\cite{mishra2018multiobjective}
   However, even with the above success of multiobjective parameter refinement including the dynamic properties, parameter fitting takes a considerable time depending on the properties in which we are interested. 
   For example, if during a particular reaction a free energy barrier 
   or infrared (IR) spectroscopy is required,
   parameter refinement is necessary for these physical properties. However, 
   this involves a considerable computational cost, and it is sometimes impossible 
   within our available computer resources; thus, parameter optimization 
   for  a nonequilibrium process is still a challenging issue.

   Thus, force field parameter refinement in ReaxFF warrants further investigation. This study focuses on how to efficiently obtain an appropriate parameter set
   for the physical properties of interest.    
   For this purpose, a new multiobjective parameter fitting process was developed.
   The distinct features in the proposed method are as follows:
   (i)  efficiently obtaining several local minima during the parameter optimization step, 
   (ii) an efficient optimization process based on an ML approach,
   and
   (iii) transferability to other parameter optimization processes given the simple structure and easy implementation. 

   In this study, the proposed ReaxFF parameter optimization method 
   is first described in detail. Then, the parameter optimization results 
   are briefly described. As a pilot test, a ReaxFF parameter force field for  Al$_2$O$_3$ is generated, and  
   the performance of the reactive MD simulation is evaluated.
   Al$_2$O$_3$ is used for many important industry materials such as     passivation films for solar cells and polishing materials for cutter grinders. 
   The Al$_2$O$_3$ crystal is usually applied as a coating in a nonequilibrium process such as    CVD or atomic layer deposition (ALD). Because it is necessary to perform the simulation    at a high temperature, reproduction of the crystal structure even at a high temperature    is important for a meaningful simulation. 
   As a pilot test of parameter optimization, 
   the MD simulation was performed for the bulk and surface crystal structure 
   to demonstrate the effectiveness of the proposed parameter-fitting approach.
   Finally, the reactive MD simulation was performed    for the pilot test.

\section{ReaxFF Parameter Optimization Using ML}
   The proposed parameter optimization scheme comprises three important steps.
   (i) The predefined initial parameters are randomly modified 
       to generate the training data set for ML (Figure~\ref{Figure01}).
  (ii) Data analysis based on ML is performed for this 
       training data set (Figure~\ref{Figure02}). 
 (iii) A grid search parameter optimization based on the ML model is performed
        (Figure~\ref{Figure03}). 
    The respective steps are denoted as 
    (i)  random parameter sampling, 
   (ii)  data analysis based on ML, and 
   (iii) grid-search parameter optimization. A detailed description of each step follows below.

    \subsection{Definition of parameter sets and random parameter sampling}
     The potential energy in ReaxFF is  
     \begin{align}
        E =&
           E_{\mathrm{bond}}    
        +  E_{\mathrm{lp}} 
        +  E_{\mathrm{over}}
        +  E_{\mathrm{under}}
      \notag \\
        +& E_{\mathrm{val}}
        +  E_{\mathrm{pen}}     
        +  E_{\mathrm{coa}} 
        +  E_{\mathrm{C2}}   
      \notag \\
        +& E_{\mathrm{tors}}  
        +  E_{\mathrm{H-bond}}
        +  E_{\mathrm{vdWaals}} 
        +  E_{\mathrm{coulomb}},
     \end{align}
where $E_{\mathrm{bond}}$ and $E_{\mathrm{lp}}$ are the potential energies 
for bond and lone pairs, respectively; 
$E_{\mathrm{over}}$ and $E_{\mathrm{under}}$ are the energy penalties for over- 
and undercoordination; 
$E_{\mathrm{val}}$ is the valence angle term; 
$E_{\mathrm{pen}}$ is the penalty energy term related to the double bond; 
$E_{\mathrm{coa}}$ is the energy of the three-body conjugation term; 
$E_{\mathrm{C2}}$ is the energy of the strong triple bond; 
$E_{\mathrm{tors}}$ is the potential energy for torsion; 
$E_{\mathrm{H-bond}}$ is the energy of hydrogen bonds; and
$E_{\mathrm{vdWaals}}$ and $E_{\mathrm{coulomb}}$ are the van der Waals 
and Coulomb interaction energies, respectively 
(for detail see \cite{van2001reaxff}).

 Thus, the ReaxFF parameter set contains the information of bonds  
 ($P_{\mathrm{be1}}$, $P_{\mathrm{be2}}$, $P_{\mathrm{bo1}}$, $P_{\mathrm{bo2}}$, etc.),
 van der Waals force ($D_{ij}$ and $R_{\mathrm{vdw}}$, etc.),
 and angle related to three-body parameters ($P_{val1}$, $P_{val2}$).
 The parameters ($P_{\mathrm{be1}}$, $P_{\mathrm{be1}}$ $\cdots$, $D_{ij}$, $R_{\mathrm{vdw}}$,
 $\cdots$ $p_{\mathrm{val2}}$ $\cdots$) are defined as a set of parameters for MD simulation,
 and these ReaxFF parameters are denoted as the parameter set.
 
\begin{figure}[h!]
          \begin{center}
             \includegraphics[clip,width=8.0cm]{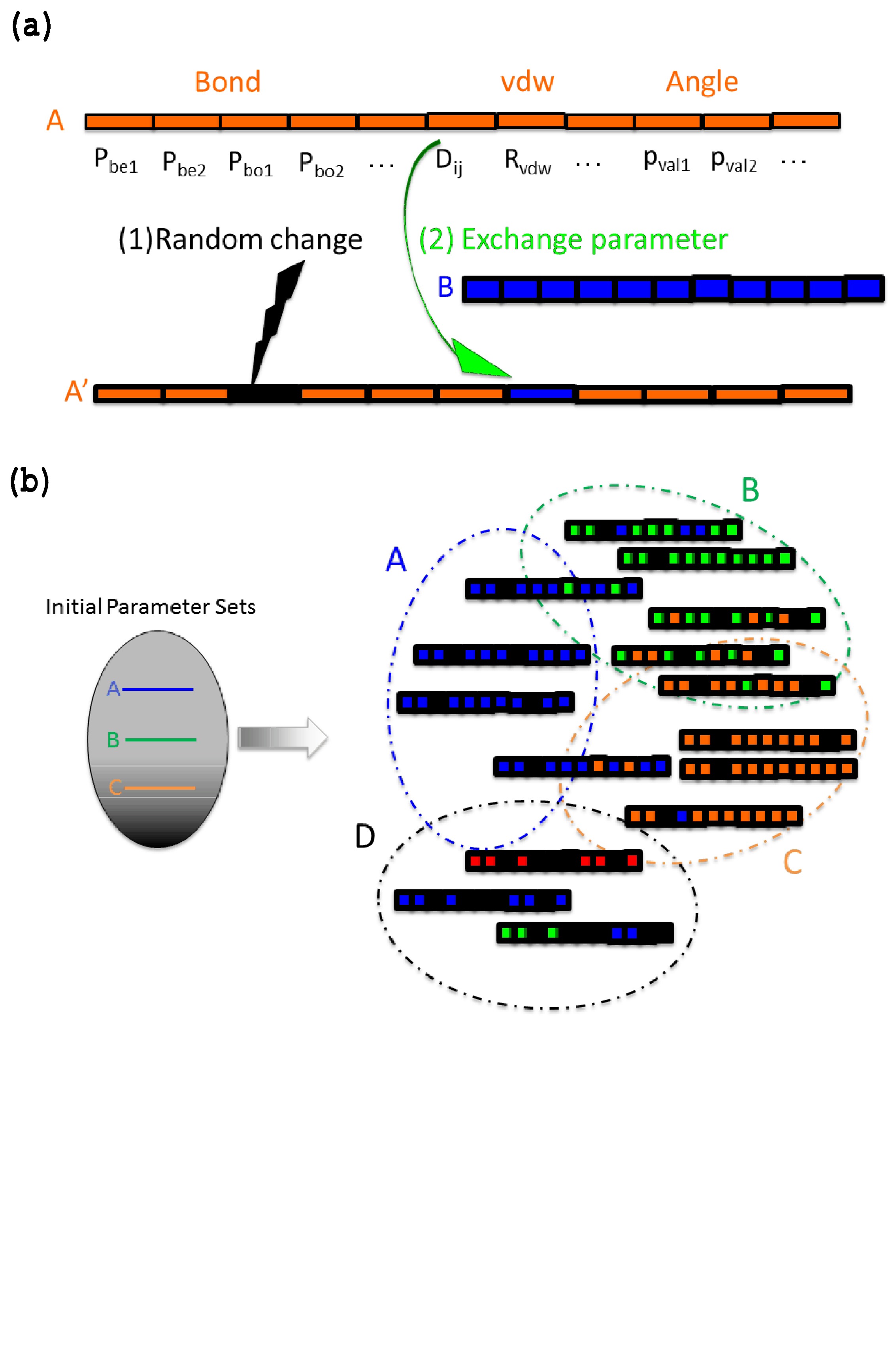} \\
          \end{center}
 	      \caption{
                 Schematic illustration for random parameter sampling.
                 (a) Definition of parameter set, and its random modification 
                     scheme.
                 (b) Entire image of the sampling space using the random modification scheme.
                 \label{Figure01}
 		      }
 	\end{figure}

 Two different types of random change are introduced to generate 
 a variety of types of parameter sets, as follows (Figure~\ref{Figure01}-(a)).
 One is a random change in the parameter set (Figure~\ref{Figure01}-(a)), 
and several parameters are selected ($\mathrm{ptype}$) randomly with a predefined probability 
 $P_\mathrm{rand}$=0.5,
 where $\mathrm{ptype}$ is the initial parameter value, such as $P_{\mathrm{be1}}$ or $P_{\mathrm{be2}}$.
 The new randomly changed parameter $\mathrm{ptype}^\prime$ is 
 \begin{align}
    \mathrm{ptype}^\prime =  \mathrm{ptype} + \delta_\mathrm{ptype}\times\mathrm{rand},
 \end{align}
 where $\delta_\mathrm{ptype}$ is 
 \begin{align}
   \delta_\mathrm{ptype} = \mathrm{C_\mathrm{scale}}\times\mathrm{ptype},
   \label{deltaparam}
 \end{align}
   where $\mathrm{C_\mathrm{scale}}$ and $\mathrm{rand}$
   are scale factor and  random number, respectively. 
   In this study, $\mathrm{C_\mathrm{scale}}=0.1$ and $-1.0 \leqq \mathrm{rand} \leqq 1.0$
   are used; therefore, the original parameter (ptype) can be changed by 10\%.
  Another type of parameter modification involved exchanging the parameter value $\mathrm{ptype}$  
  for a value in another initial parameter set with a probability $P_\mathrm{exch}$, and 0.2 is used 
  for the probability of exchange parameter sets in this study.
  
  This modification in the parameter set is inspired by the genetic algorithm\cite{reaxFFGA01}
  and efficient sampling of various kinds of parameter sets is possible. 
  As an example, a schematic illustration of parameter sampling 
  is shown in Figure~\ref{Figure01}-(b).
  In Figure~\ref{Figure01}-(b), the different colors denote the different kinds 
  of initial parameter sets, and three initial parameter sets are 
  shown in Figure~\ref{Figure01}-(b).
  By performing the random sampling according to Figure~\ref{Figure01}-(a),
  the parameter set spans over the space in the initial parameter sets 
  of A, B, and C, as shown on the left-hand side of Figure~\ref{Figure01}-(b).
  In addition, a new sampling space is sometimes found by the random change
  of the initial parameter sets of D, as shown on the right-hand side of Figure~\ref{Figure01}-(b); thereby, efficient parameter sampling becomes possible.

    \subsection{Evaluation of  the ReaxFF parameter sets}
    The previous section introduced the random parameter sampling 
    to obtain the training data set (ReaxFF parameter sets).
    For the data analysis based on ML,
    each training data set (ReaxFF parameter set) should have a score 
    that represents the validity of the data set.

    The score $S(p_i)$ for each training ReaxFF parameter set $p_i$  can 
    be evaluated as follows:
    \begin{align}
       S(p_i)       
     = & 
       \sum^{N_{\mathrm{QMtype}}}_j 
       \frac{w_j S_{j}(p_i)}
            {N_{\mathrm{QMtype}}}
       \label{ScoreTot}
     \\
       S_{j}(p_i) 
       \label{ScoreLocal}
     = & 
       \sqrt{
         \sum^{N_j^{\mathrm{QMtype}}}_k
         \frac{
          \left(
            Q^\mathrm{ReaxFF}_{k,j}(p_i)
        -
            Q^\mathrm{QM}_{k,j}
          \right)^2
         }{
          N_j^{\mathrm{QM}}
         }
       },
    \end{align}
    where $N_{\mathrm{QMtype}}$ is the number of geometry sets,
    and each geometry set $j$ contains $N_j^{\mathrm{QMtype}}$ 
    different structures that represent a particular potential energy curve 
    or physical properties such as the potential energy along 
    the volume change in the $\alpha$-Al$_2$O$_3$ crystal or the bond cleavage of H$_2$O. 
    $S_{j}(p_i)$ and $w_j$ are the score and weight for each geometry set $j$,
    and we used the root mean square error (rmse) to evaluate  $S_{j}(p_i)$.
    $Q^\mathrm{ReaxFF}_{k,j}(p_i)$ and  $Q^\mathrm{QM}_{k,j}$
    are quantities evaluated with ReaxFF and QM for each $k$th structure 
    in the geometrical set $j$.
    It is possible to use any kind of physical properties for the quantities
    $Q^\mathrm{ReaxFF}_{k,j}(p_i)$ and  $Q^\mathrm{QM}_{k,j}$,
    and we used the potential energy in this study.

    Because the absolute value of $Q^\mathrm{QM}_{k,j}$ is 
    significantly different between the respective geometry sets $j$,
    $Q^\mathrm{ReaxFF}_{k,j}(p_i)$ and  $Q^\mathrm{QM}_{k,j}$ were standardized:
   \begin{align}
      \label{normalizeScore}
      Q^{\mathrm{ReaxFF},\prime}_{k,j}(p_i) = \frac{Q^\mathrm{ReaxFF}_{k,j}(p_i) - \mu^j}{\sigma^j},
    \\
      Q^{\mathrm{QM},\prime}_{k,j}          = \frac{Q^\mathrm{QM}_{k,j}          - \mu^j}{\sigma^j},
   \end{align}
    where $\mu^j$ and $\sigma^j$ are the average and standard deviation (SD) for the geometry set $j$
    of  $Q^\mathrm{QM}_{k,j}$.
    The standardized quantities $Q^{\mathrm{ReaxFF},\prime}_{k,j}(p_i)$ are used in eq~\ref{ScoreLocal} 
    instead of  the original values  $Q^\mathrm{ReaxFF}_{k,j}(p_i)$.

    \subsection{Data analysis based on ML}
\begin{figure}[h!]
          \begin{center}
             \includegraphics[clip,width=6.0cm]{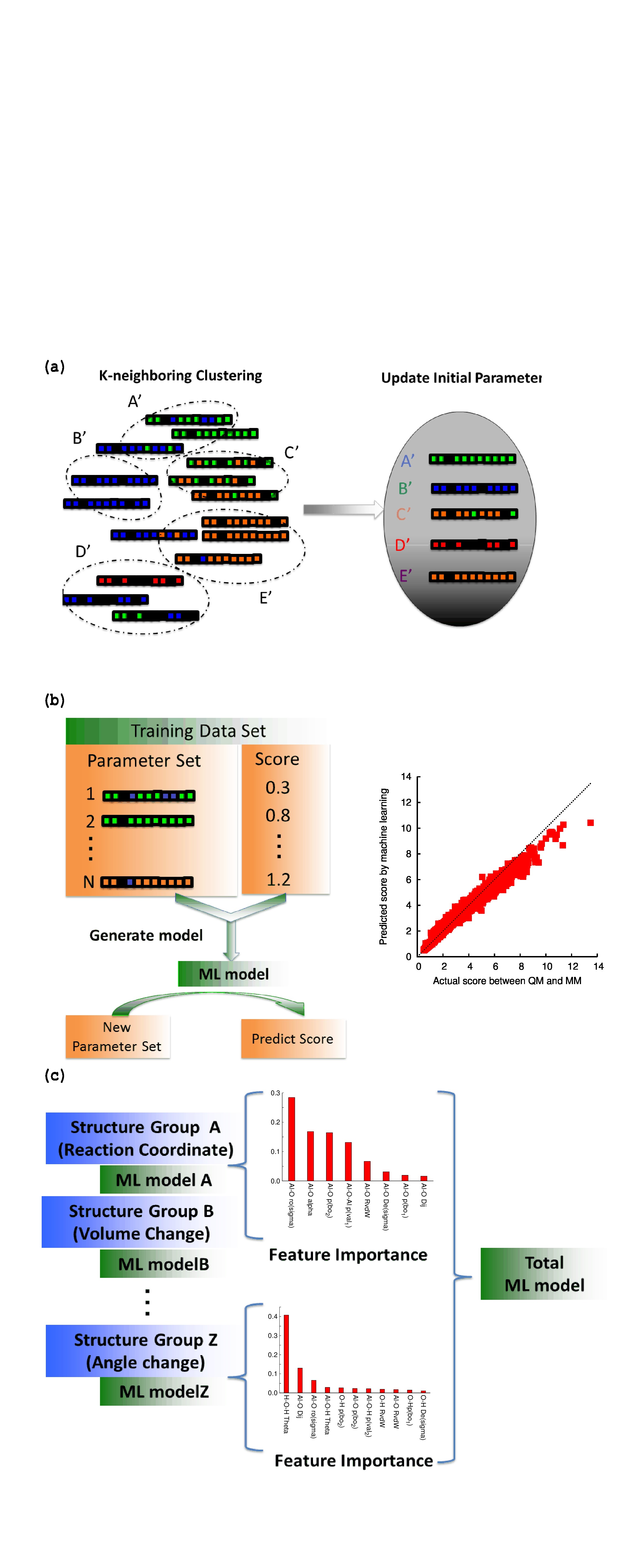} \\
          \end{center}
 	      \caption{
                 Three important ML methods used in this study.
                 (a) K-nearest neighbor clustering to select initial ReaxFF parameter set.
                 (b) Generate an ML model to predict the score.
                     The definition of score is described in eq~\ref{ScoreTot}.
                     On the right: the horizontal axis is the total score $S(p_i)$ estimated  
                     by the ReaxFF parameter force field, and the vertical axis is the total score
                     predicted by the ML model.
                 (c) Schematic illustration for the entire ML model 
                     and the feature importance for each structure group. 
                 \label{Figure02}
 		      }
 	\end{figure}

    Now the training data set contains 
    the parameter set $p_i$, 
    the total score $S(p_i)$,
    and the scores $S_{j}(p_i)$ for each geometry set $j$ (for each different kind of property).
    By using this training data set, 
    three different kinds of data analysis were performed
    based on the ML approach:
    (a) update the initial parameter set by the $k$-nearest neighbor 
        algorithm (Figure~\ref{Figure02}-(a)),
    (b) generate the ML model (Figure~\ref{Figure02}-(b)),
        and 
    (c) extract the feature importance (Figure~\ref{Figure02}-(c)).
  
     In step (a), the initial parameter set is updated by the $k$-nearest neighbor algorithm 
     (Figure~\ref{Figure02}-(a)). As shown in  Figure~\ref{Figure02}-(a), the
     $k$-nearest neighbor algorithm separates the training data sets 
     into $k$ groups, 
     and the $k$-nearest neighbor algorithm 
     makes the distance between the groups as large as possible.
     In this study, eight was used for $k$, and increasing $k$
     improves the convergence of parameter optimization,  
     but the computational time increases because of the following grid 
     search optimization (see next section).
     Thus, the number of $k$ between six and eight seems an appropriate choice for parameter $k$.
     From each group classified by the $k$-nearest neighbor algorithm, 
     the parameter set that has the lowest total score 
     is selected, and the initial parameter sets are updated as 
     shown in Figure~\ref{Figure02}-(a).
     
     Figure~\ref{Figure02}-(b) is a schematic 
     illustration of the ML approach.
     The ML model was constructed so as to
     reproduce the score ($S(p_i)$ or $S_{j}(p_i)$) from the parameter sets ($p_i$)
     (see the right-hand side of Figure~\ref{Figure02}-(b)).
     In this study, the training data set contained
     the total score $S(p_i)$, and $S_{j}(p_i)$ for each geometry set $j$;
     therefore, for $N_{\mathrm{QMtype}}$ + 1, an independent ML model 
     can be constructed. To make ML models,
     a random forest regressor model was used in this study.
     One of the merits of using the random forest regressor is the feature importance of the by-product,      
     which describes the sensitivity of respective elements in the parameter set.

     The entire structure of the ML model is depicted in 
     Figure~\ref{Figure02}-(c). The training data set contains a score ( $S_{j}(p_i)$)  
     for each  geometry set $j$ ($N_{\mathrm{QMtype}}$ different physical properties), 
     and each geometry set has an ML model and feature importance.  
     It is also possible to construct an ML model for the total score ($S(p_i)$).
      The grid search parameter optimization was performed using the ML model for the total score and each geometry set, as described below.

    \subsection{Grid search parameter optimization}
\begin{figure}[h!]
          \begin{center}
             \includegraphics[clip,width=8.0cm]{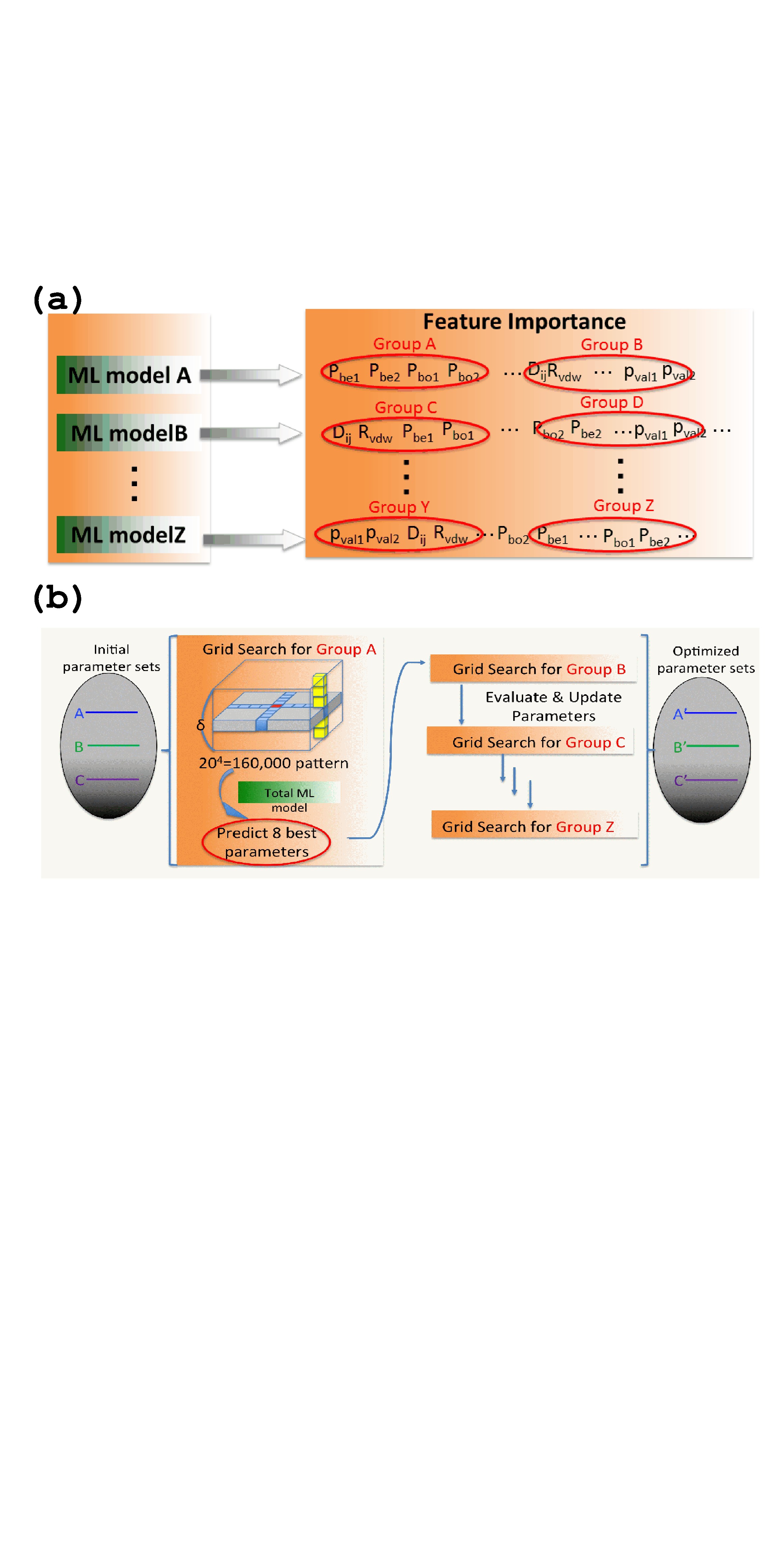} \\
          \end{center}
 	      \caption{
             Schematic illustration for grid search optimization.
             ML model A denotes the ML model for structure group A,
             and $\delta$ in (b) corresponds to $\delta$ 
             in eq~\ref{deltaparam}.
             \label{Figure03}
 		  }
 	\end{figure}

     The reaming process for ReaxFF parameter optimization was the grid search 
     parameter optimization based on the ML model generated 
     as described in the previous section.
     The schematic illustration for grid search parameter optimization is shown 
     in Figure~\ref{Figure03}.

     Each geometry set ($N_{\mathrm{QMtype}}$ different physical properties) contains 
     the independent ML model, and distinct types of feature importance.
     According to the feature importance, the parameter set is split into several groups,
     and each group contains four parameters.
     For example in Figure~\ref{Figure03}-(a), ML model A contains 
     the parameters 
     ($P_{be1}$, $P_{be2}$, $P_{bo1}$, $P_{bo2}$, $\cdots$,
     $D_{ij}$, $R_{vdw}$, $p_{val1}$, $p_{val2}$, $\cdots$).
     The four parameters ($P_{be1}$, $P_{be2}$, $P_{bo1}$, $P_{bo2}$) 
     are the most important parameters for geometry set A,
     and this four-parameter set was denoted as  Group A.
     Likewise, the four less important parameters such as ($D_{ij}$, $R_{vdw}$, $p_{val1}$, $p_{val2}$) 
     were denoted as group B. Because each ML model contains distinct feature importance levels,
     the respective groups (A, B, $\cdots$, Z) contain different combinations of parameters.
     
     Then the grid search parameter optimization is performed for each group (A, B, $\cdots$, Z) 
     independently, and the summary of the grid search parameter optimization is 
     shown in Figure~\ref{Figure03}-(b).
     First, the initial preliminary parameter sets were prepared according to Figure~\ref{Figure02}-(a). 
     To perform grid search parameter optimization, the initial parameters 
     span $\delta_\mathrm{ptype}$ around their initial values,
     and the range $\delta_\mathrm{ptype}$ is split into 20 grid points.
     As each group contains four parameters, the total number of grid points is 
     $20^4=160,000$, and the direct evaluation of the total score in eq~\ref{ScoreTot}  
     is time-consuming. The ML model can predict the total score within negligible 
     computational time; therefore, the computational time can be significantly reduced.
     
     Using the ML model, eight best parameter sets were predicted
     from the 160,000 different parameter sets, 
     and the total scores were evaluated only for these eight parameter sets
     using eq~\ref{ScoreTot}. 
     If the evaluated total score is less than the score in the initial parameter set,
     the initial parameter set is updated.
     The grid search parameter optimization procedure is iteratively performed 
     for each group A, B, $\cdots$, Z  (Figure~\ref{Figure03}).

     \subsection{Summary of parameter optimization based on the ML approach}
    
       The previous sections described the respective parts of parameter 
       optimization; the purpose of this section is to connect the respective steps to 
       a cycle of parameter optimization.
       
       The random parameter sampling illustrated in Figure~\ref{Figure01} plays 
       two important roles.
       The random sampling can escape trapping a local minimum during parameter optimization, 
       and could efficiently sample the training data set for the ML model.

       The ML model was constructed based on the training data set, and involves
       three important roles.
       First, the $k$-nearest neighbor algorithm selects each initial parameter set as far as possible,
       providing efficient finding of $k$ possible local minima. 
       Second, the random forest regressor model can efficiently predict the parameter set
       containing the lowest total score, and significantly reduce the computational cost 
       by avoiding the evaluation of many redundant parameter sets. 
       Third, the feature importance estimated by the random forest regressor  
       provides a guide for the optimization of the respective parameters, and the groupwise grid search can        be effective.

       By iteratively performing random structure sampling, ML analysis,
       and grid search parameter optimization, we can obtain $k$ possible local minima, which contain different uncertainties for       the other physical properties, such as crystal structure at high temperature. 
       In the following discussion, the effectiveness of this approach 
       for performing an MD simulation is demonstrated.

\section{Computational Detail}
   The computational models used for potential energy evaluated by QM were  
   based on the crystal structure of Al$_2$O$_3$.
   The geometry sets for the quantity evaluation in eq~\ref{ScoreLocal}
   were the potential energy curve along the volume change in $\alpha$-Al$_2$O$_3$ 
   crystal structure,
   the volume change in $\gamma$-Al$_2$O$_3$ crystal structure, the bond cleavage in H$_2$O, HCl, 
   and AlCl$_3$,
   the angle potential energy curve for H$_2$O, and the absorption energy for H$_2$O
   on the (0001) and (11$\overline{2}$0) surface of $\alpha$-Al$_2$O$_3$.

   A plane wave-based density functional theory (DFT) program Castep was used\cite{CASTEP1,CASTEP2} 
   with the Perdew, Burke and Ernzerhof (PBE) functional.\cite{PBE1,PBE2}
   All the calculations were performed with ultrasoft core potentials\cite{ultrasoft} generated on the fly, 
   and the cutoff energy was set to 571.4 eV. 
   The electronic configurations of the atoms were  
   Al:3s$^2$ 3p$^1$, 
   O:2s$^2$ 2p$^4$, 
   H:1s$^1$, and
   Cl:3s$^2$ 3p$^5$. 
   We adopted the convergence criterion of 0.03 eV/A for geometry optimizations, 
   and 0.0001 eV for the self-consistent field calculations of electronic states.
  A 2 $\times$ 2 $\times$ 2 uniform  mesh  for k-space integrations was used 
   for all the bulk simulations, and a 2 $\times$ 2 $\times$ 1 mesh was used for the 
   surface models.

   To evaluate the validity of the obtained ReaxFF parameter set,
   MD simulations were  performed 
   for the crystal structure of $\alpha$-Al$_2$O$_3$.
   The system size was 42.8 $\times$ 49.45 $\times$ 38.973 and the total number of atoms was 9720.
   The coordination number (CN) of Al atoms and the angle between O--Al--O  were  evaluated regardless of
   whether or not the crystal structure was maintained during the MD 
   simulation even in the relatively high-temperature region (2000 K).

   To show the feasibility of the optimized ReaxFF parameter set,
   the MD simulations were also performed with surface structures.
   Two kinds of surface structure, (0001) and (11$\overline{2}$0), were evaluated. 
   A previous study of thermodynamic simulations\cite{Al2O3Firstprinciple,CVDAl} suggests 
   that the $\alpha$-Al$_2$O$_3$ surface is terminated by Cl; therefore,
   the simulation models were also terminated by Cl, 
   and the crystal structure was investigated.
   The system size for the surface model  was 32.97 $\times$ 33.313 $\times$ 60.625, 
   and the total number of atoms was 7952. For the pilot test of the ReaxFF parameter,
   a reactive MD simulation was performed 
   using the surface model of $\alpha$-Al$_2$O$_3$, and the 
   HCl evolution reaction was evaluated.

   All the MD simulations in bulk structures were performed at  2000 K, 
   which nearly corresponds to the melting point of the $\alpha$-Al$_2$O$_3$ crystal.
   The MD simulations on the surface model were performed 
   at 1223 K, because the CVD experiments on $\alpha$-Al$_2$O$_3$ are typically conducted 
   at 1223 K.
   For the MD simulation of bulk and surface structures,
   NVT MD simulation was performed 
   with a Nose--Hoover thermostat, and the velocity Verlet integrator was 
   used for time integration, and the time step was 0.1 fs.
   The number of time steps was 100,000, which corresponds to the 10 ps simulation time.

\section{Results and Discussion}

\subsection{Parameter fitting}

  The rmse and maximum errors (MaxErr) between the ReaxFF parameter 
  and the QM potential energy are summarized in Table~\ref{TABLE01}.
  The rmse and MaxErr were normalized by the SD
  (see eq~\ref{normalizeScore}).
  As shown in  Table~\ref{TABLE01}, all the rmse values estimated by the
  ReaxFF parameter set  are around 0.1, and the parameters could reproduce the 
  QM result within 3$\sim$5\% errors. 
  Thus, almost all the optimized parameter sets represent a reasonable 
  agreement with the QM calculation (see below for more detail 
  in the comparison of the potential energy between QM and the ReaxFF parameter).

  The point of the proposed approach is that the single ReaxFF parameter optimization task 
  generates $k$ possible local minima; therefore,
  $k$ independent ReaxFF parameter sets could be obtained directly.
  To show the difference between the optimized parameter sets,
  a summary of the respective parameter sets is shown in Table~\ref{TABLE03}.
  A significant difference is observed between the parameter sets 
  in terms of $\alpha$-Al$_2$O$_3$ crystal structure;
  the parameters related to Al and O are shown in Table~\ref{TABLE03}. 
  As shown in Table~\ref{TABLE03}, the differences between the parameters 
  are relatively large, especially the angle-related parameters p(val$_1$) 
  and p(val$_2$), which differ by around 24\% (std/avg) and 17\% (std/avg), respectively. 
  It is interesting to obtain these different kinds of ReaxFF parameters  
  with the power of the ML technique.

\begin{figure}[h!]
          \begin{center}
             \includegraphics[clip,width=8.0cm]{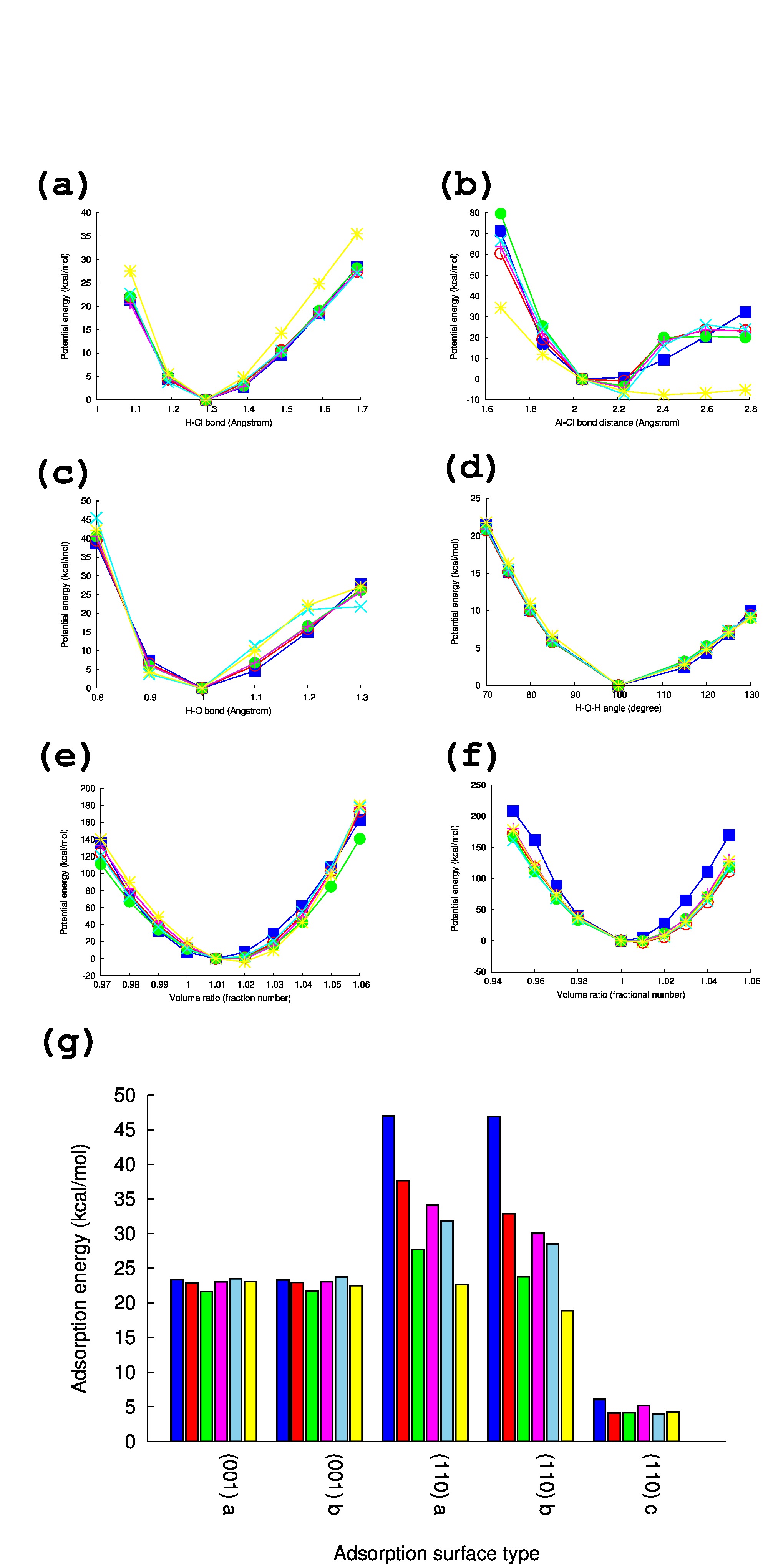} \\
          \end{center}
 	      \caption{
             Potential energy curve estimated by respective ReaxFF parameters and QM.
             The blue filled squares are the potential energy estimated by QM.
             The red, green, magenta, sky-blue, and yellow colors are 
             the potential energy estimated by ReaxFF parameter sets A--E,
             respectively.
             (a) H--Cl bond, 
             (b) Al--Cl bond in AlCl$_3$, 
             (c) H--O bond in water,  
             (d) H--O--H angle in water, 
             (e) energy volume curve for  $\alpha$-Al$_2$O$_3$ crystal,
             (f) energy volume curve for  $\gamma$-Al$_2$O$_3$ crystal,
             and
             (g) H$_2$O adsorption energy for (0001) and (11$\overline{2}$0) surface of  $\alpha$-Al$_2$O$_3$.
                 (001)a denotes adsorption of water with weak H bonding.
                 (001)b denotes adsorption of water without  weak H bonding. 
                 (110)a denotes adsorption of water, and this water is split into Al--OH and H--OAl.
                        Then the H atoms have very weak H bonds.
                 (110)b is similar to (110)a without the H bond.
                 (110)c is only H bonding between H and O, but no covalent bond between
                        Al and O of water.
             \label{Figure04}
 		  }
 	\end{figure}

  The parameter fitting results (A$\sim$H in  Table~\ref{TABLE01}) 
  in terms of the respective 
  structure data sets are shown in Figure~\ref{Figure04} for A$\sim$E, 
  and the Supporting Information (SI) for F$\sim$H. 
  (The MD simulation results with parameter sets F$\sim$H 
   present similar results to A$\sim$E, and the results are shown in the 
   SI.)
  The comparison between the QM calculations (blue filled squares) 
  and the ReaxFF parameter results shows good agreement with each other.
  The extent of errors shown in Figure~\ref{Figure04} is often found
  in other ReaxFF parameter optimizations,\cite{ReaxFFQMvsMM01,ReaxFFQMvsMM02,ReaxFFQMvsMM03,ReaxFFQMvsMM04,ReaxFFQMvsMM05,ReaxFFQMvsMM06,ReaxFFQMvsMM07}
  which suggests that the optimized parameter set was reasonable 
  for reactive MD simulation.
  The next section presents how the parameter sets A$\sim$H
  affect the crystal structures during the MD simulations
  at high temperature (2000 K).

\subsection{Evaluation of the parameter sets using MD in $\alpha$-Al$_2$O$_3$}

  To show a specific example for the difference in physical property depending 
  on the parameter sets A$\sim$H,
  MD simulations were performed with 
  $\alpha$-Al$_2$O$_3$ for each parameter set,
  and we investigated whether the crystal structure of $\alpha$-Al$_2$O$_3$ was maintained or not.
  For this purpose, the CN of Al was calculated.
  The Al atom possesses six bonds with O, therefore
  the CN is six. If the crystal structure of $\alpha$-Al$_2$O$_3$  
  is maintained, all the CNs of Al should be six, 
  so that the ratio of CN six of Al is evaluated 
  to determine the crystalline structure during the MD simulation.
  For simplicity, we denote the ratio of CN 6 of Al 
  as the crystal structure ratio.

  The results of MD simulation using the respective optimized 
  parameter set are summarized in Table~\ref{TABLE02}.
  As shown in Table~\ref{TABLE02}, 
  a significant deviation of the crystal structure ratio is observed 
  between the parameter sets A$\sim$H from 25\% to 89\%.
  The crystal structure of $\alpha$-Al$_2$O$_3$  
  is maintained for about 90\%
  in the MD simulation using with parameter set A. 
  In contrast, with parameter set E, 
  most of the Al atoms deviate significantly, and 
  the crystal structure ratio is only 25.87\%.
  
\begin{figure}[h!]
          \begin{center}
             \includegraphics[clip,width=8.0cm]{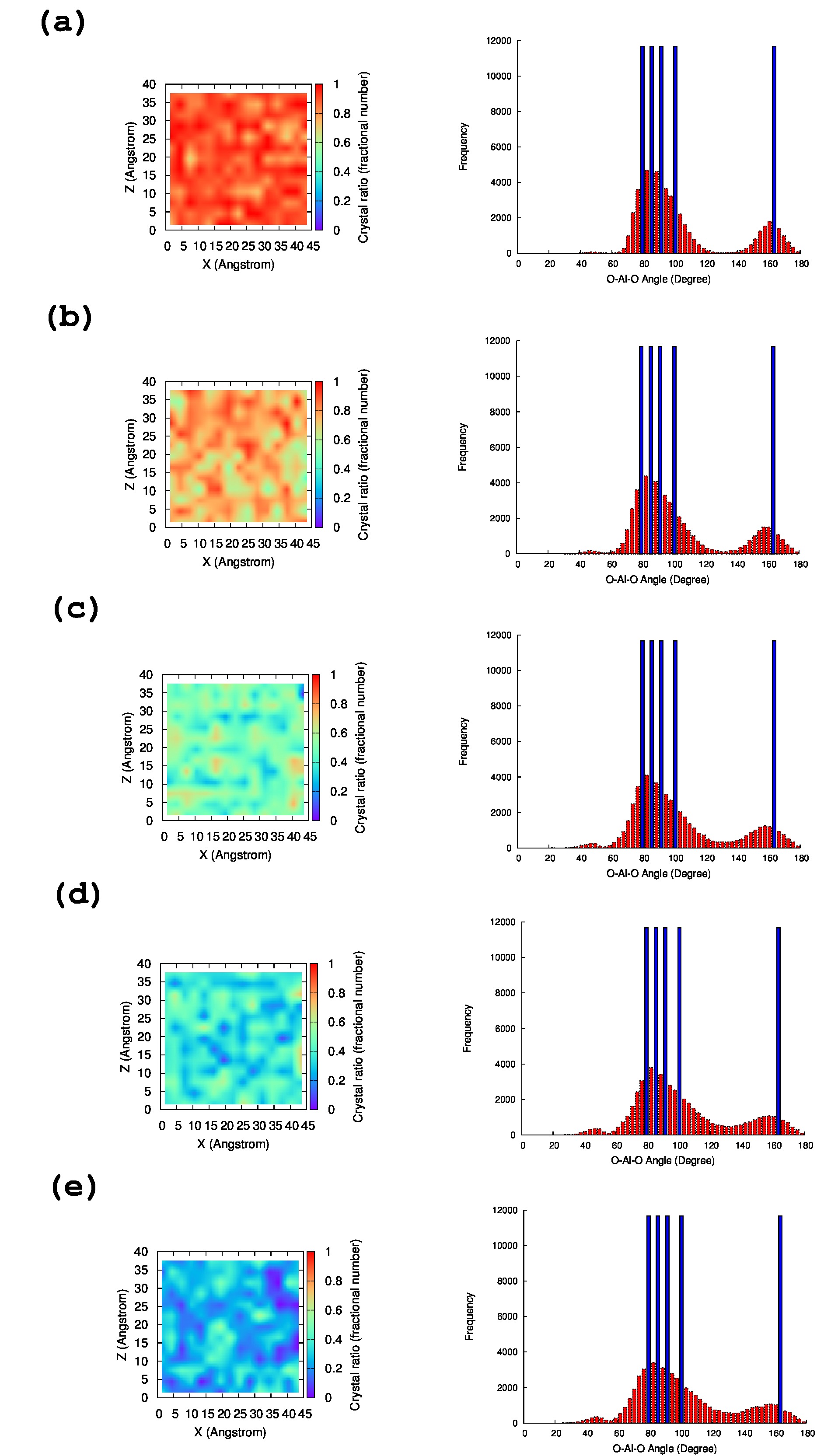} \\
          \end{center}
 	      \caption{
              Structure analysis for $\alpha$-Al$_2$O$_3$  during the MD simulation at 2000 K.
             (left): Crystal structure analysis  for the respective ReaxFF parameter sets.  
                     X and Z denote the Cartesian coordinates, and the color bar is the 
                     crystal ratio (see main manuscript for detailed definition).
             (right): Histogram for the O--Al--O angle; the horizontal axis is angle (degree) 
                      and the vertical axis is frequency.
             (a)--(e) are the simulation results using parameter sets A--E, respectively.
             \label{Figure05}
 		  }
 	\end{figure}

  To investigate further the  crystal structure ratio at 2000 K,
  Figure~\ref{Figure05} presents a graphic illustration of the crystal structure ratio, 
  and the O--Al--O angle using the parameter sets A$\sim$E (see the SI for F and G).
  The results using parameter set A mostly maintain the crystal structure, 
  and the red color in  Figure~\ref{Figure05} sharply decreases from parameter sets A to E.
  Instead, the blue color increases systematically from A to E, which denotes that
  the geometry obtained by parameter set E becomes amorphous.
  Likewise, the angle between O--Al--O broadens toward parameter sets A to E. 
  For parameter set A, the angle is mostly around the crystal structure at 0 K,  
  but the distribution of the angle deviates significantly in the MD simulation
  using parameter set E.
  With the analysis of CN and the distribution of the O--Al--O angle,
  we can conclude that parameter set A is the best choice for performing 
  the reactive MD simulation.

  As shown in the above example of parameter sets,   
  the well-fitting parameters (e.g., parameter set F) are sometimes not an appropriate 
  choice, because the training structure data set always contains an uncertainty
  for a particular physical property (in this case, the crystal ratio at 2000 K).
  One of the possible solutions is increasing the number of training structure data sets;
  then, the global minimum should be the reasonable parameter set.
  However, preparing an adequate training structure data set requires expert 
  experience, and it is usually difficult to know how many training structure data sets are 
  actually necessary.
  Therefore, this approach to find only one global minimum of a ReaxFF parameter 
  requires  iterative trial and error to perform a reactive MD
  simulation. 

  The approach in this study can reduce the complex procedure for parameter 
  optimization by trying to find $k$ different local minima.
  The $k$-nearest neighbor algorithm separates the data set as far as possible.
  Therefore, the local minima differ from each other. 
  Consequently, the parameter sets likely contain different kinds of 
  uncertainty for physical properties (Figure~\ref{Figure05}). 
  In this case, parameter set A provides excellent performance 
  for reproducing the crystal structure at 2000 K.

\subsection{Reactive MD simulation on the $\alpha$-Al$_2$O$_3$ surface}

   The reactive MD simulation was performed using the optimized parameter set A.
   The CVD experiment and theoretical thermodynamic study reported that 
   the surface structure of  $\alpha$-Al$_2$O$_3$ is terminated by Cl\cite{Al2O3Firstprinciple,CVDAl}
   and the HCl evolution step is an important process in the CVD experiment.
   Therefore, the HCl evolution reaction was investigated in this study.

\begin{figure}[h!]
          \begin{center}
             \includegraphics[clip,width=14.0cm]{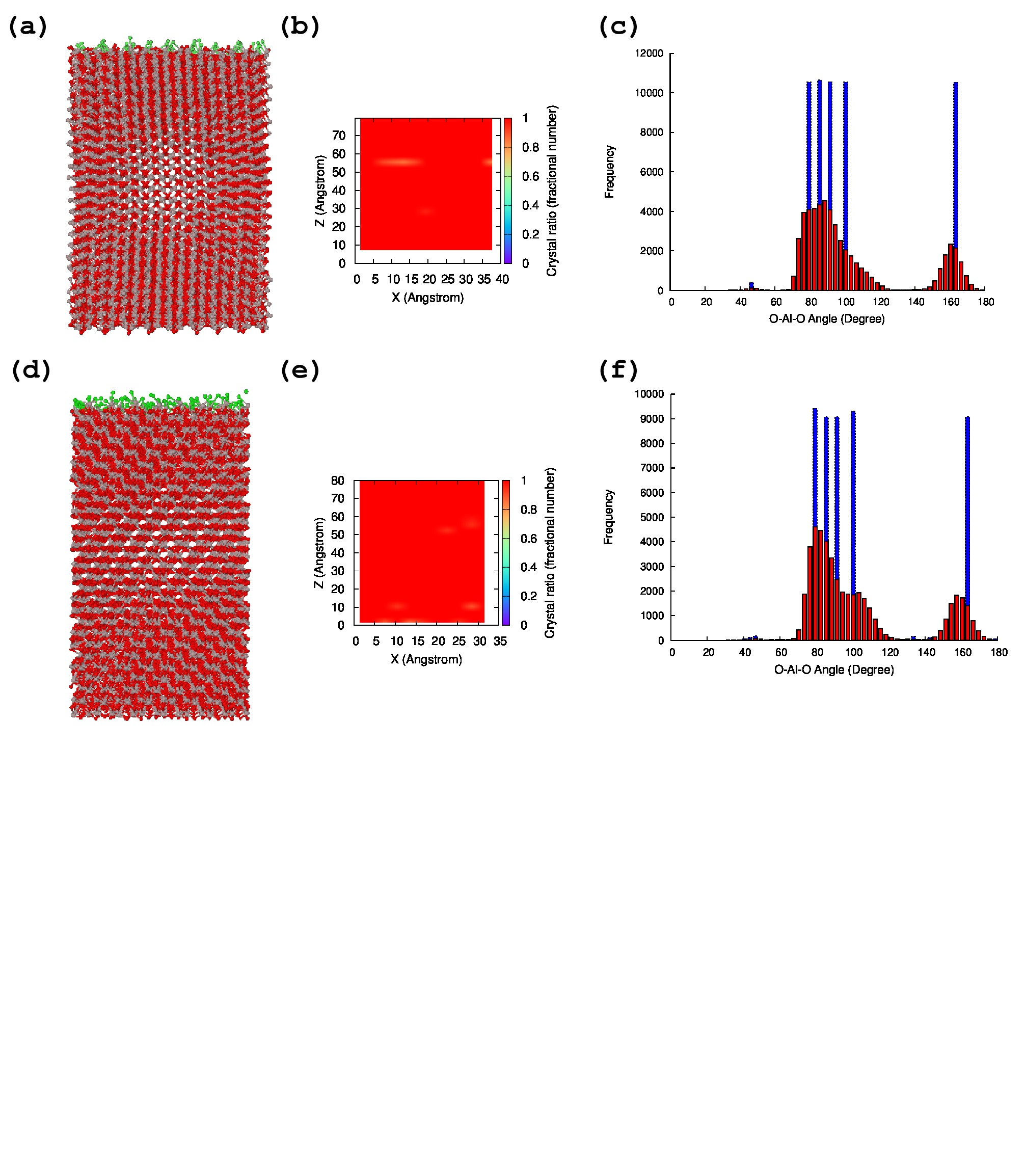} \\
          \end{center}
 	      \caption{
             Surface structure analysis $\alpha$-Al$_2$O$_3$ during the MD simulation at 1223K.
             (a)--(c), respectively, are the analysis for the (11$\overline{2}$0) surface, and
             (d)--(f) are  the analysis for  the (0001) surface.
             (a) and (d) are  the side views of the simulation model.
                 (Red denotes O atoms, magenta is Al atoms, 
                 and green is Cl.)
             (b) and (e) are the crystal ratios (for detail see the caption in Figure~\ref{Figure05}).
             (c) and (f) are  the histograms for angle distribution. 
             \label{Figure06}
 		  }
 	\end{figure}

   For this purpose, the surface structures of $\alpha$-Al$_2$O$_3$ (0001) and (11$\overline{2}$0)
   were investigated using the optimized parameter set A.
   The result of the MD simulation of the surface structure is shown in  Figure~\ref{Figure06}.
   In Figure~\ref{Figure06},  two different kinds of surface structure are shown.
   The surface structures of  $\alpha$-Al$_2$O$_3$ (0001) are shown in Figure~\ref{Figure06}-(a)--(c),
   and the surface structures of $\alpha$-Al$_2$O$_3$ (11$\overline{2}$0) are 
   shown in Figure~\ref{Figure06}-(d)--(f).
   As shown in Figure~\ref{Figure06}, the crystal structure is reasonably maintained during
   the 10 ps MD simulation, and most of the Al atoms have 
   six coordination between O atoms. In addition, the angle O--Al--O shows reasonable agreement    with the experimental crystal structure. Therefore, MD simulation 
   on the surface structure is possible. 

\begin{figure}[h!]
          \begin{center}
             \includegraphics[width=8.0cm]{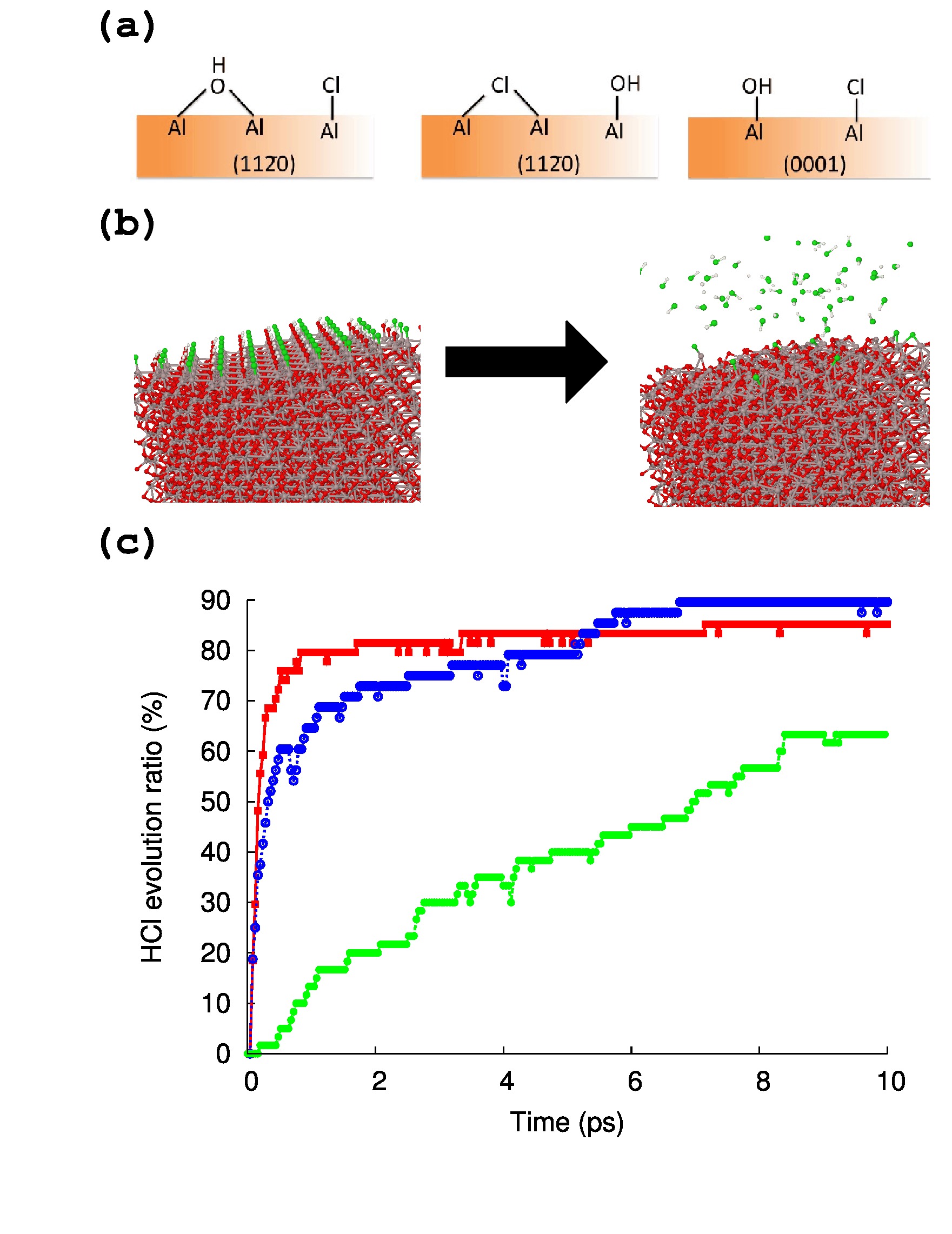} \\
          \end{center}
 	      \caption{
             Analysis for reactive MD simulation
             for the surface structure of $\alpha$-Al$_2$O$_3$.
             (a) Schematic illustration for the initial surface model  
                 for (11$\overline{2}$0) and (0001).
             (b) Initial and end of surface structure for (0001).
             (c) HCl evolution during the MD simulation.
             \label{Figure07}
 		  }
 	\end{figure}

   As the MD simulation on the surface is possible using parameter set A, the
   HCl evolution reaction was investigated by reactive MD simulation.  
   For this purpose, surface models for both  (0001) and (11$\overline{2}$0) were prepared
   as shown in Figure~\ref{Figure07}-(a). 
   On the (0001) surface, there is only one type of site, and each site is occupied by Cl or OH 
   (this model is denoted as (0001)).
   This model contains excessive OH compared with the experiment to accelerate the 
   HCl evolution reaction.
   Likewise, the (11$\overline{2}$0) surface model is shown in  Figure~\ref{Figure07}-(a), 
   and two different site types are observed on the (11$\overline{2}$0)  surface.
   Thus, two different types of surface structures were evaluated.
   One is the model that contains OH bridged between two Al atoms, and Cl attached on one Al atom    (Figure~\ref{Figure07}-(a), this model is denoted as (11$\overline{2}$0)-(a)).
   The other is a model that contains Cl bridged between two Al atoms, and OH attached on one Al atom    (this model is denoted as (11$\overline{2}$0)-(b)). 
   In total, three independent reactive MD simulations ((001), (11$\overline{2}$0)-(a), and (11$\overline{2}$0)-(b))    were performed. 
   
   The initial and final structures after the 10 ps simulation are shown in  Figure~\ref{Figure07}-(b),
   and the HCl evolution reactions during the MD simulations are depicted in  Figure~\ref{Figure07}-(c).
   Significant differences are observed between the three  surface models.
   Most of the HCl molecules immediately evolve at 1223 K in the simulation model (0001) and (11$\overline{2}$0)-(a), 
   whereas the evolution rate of HCl in (11$\overline{2}$0)-(b) is significantly slower than the others.
   This slow HCl reaction rate in  (11$\overline{2}$0)-(b) is caused by the tightly binding Cl atoms 
   on the two Al atoms; the final HCl evolution reaction ratio on  (11$\overline{2}$0)-(b) is only 63.33\%.
   In contrast, most of the HCl molecules evolve in the (0001) model in the initial 3 ps, but the HCl evolution    reaction on the (0001) surface stops at 85.19\% (HCl:ClAl = 8:2), whereas the HCl evolution on the (11$\overline{2}$0)-(a) 
   surface continues for 10 ps, and the final HCl evolution ratio is 89.58\%. 
   
   The rate-determining step seems different between each surface model. 
   In the initial steps, because most of the H is located very near Cl atoms,
   the reaction rate is determined by the HCl evolution from Al--OH and Al--Cl.
   The reaction rate of HCl evolution at the surface (11$\overline{2}$0)-(b) 
   is significantly slower than in the other surface model, and this HCl reaction 
   seems to be the rate-determining step.
   After the HCl evolution reaction occurred, the OH is seldom located next to Cl, 
   and the rate-determining step is the H transfer between the O sites.
   This conclusion is quite similar to the previously reported theoretical study\cite{Al2O3Firstprinciple,CVDAl}
   and it seems that our reactive MD simulation is reasonable for these three different surface models.

\subsection{Conclusion}
   A new method to optimize the reactive force field has been developed 
   based on ML. 
   Three important steps were introduced in this study:  the $k$-nearest neighbor algorithm to locate the possible local minima,    a random forest regressor to construct the ML model, 
   and grid search optimization to optimize the parameter set based on the ML model. 

   Using the ML technique,
   several optimized parameter sets 
   that differ as far as possible could be obtained.
   All the obtained parameter sets reasonably reproduce the potential energy estimated by QM,
   but the physical properties are very different from each other, because the parameter optimization 
   inevitably contains uncertainty for particular physical properties.

   By evaluating the CN of Al  and the  O--Al--O angle,
   it is possible to know which parameter set is an appropriate choice for performing 
   the MD simulation at high temperature.
   Then, as a pilot test, a reactive MD simulation was 
   performed  to analyze the HCl evolution reaction at the surface of  $\alpha$-Al$_2$O$_3$.
   To this end, three different types of Cl--Al binding models were investigated, 
   and the HCl evolution from Cl bridged between two Al atoms resulted in 
   a significantly slower reaction rate than in other models.
   Such different types of Cl binding models could be reasonably evaluated 
   by the parameter obtained in this study.

   The proposed optimization process could simplify the complex process of 
   ReaxFF parameter optimization with reasonable computer resources,
   which suggests that this strategy could be applied to simulate many 
   reactive MD simulations in material science.

\subsection{ACKNOWLEDGMENTS}
This research used the computational resources of the Supercomputer system ITO 
in R.I.I.T at Kyushu University as a national joint-usage/research center.
We thank Professor Momoji Kubo of Institute for Materials Research, 
Tohoku University for the helpful discussions about the reactive MD simulation.

\bibliographystyle{aip}
\bibliography{reaxff.parmopt}
   
\newpage

TABLE captions.

\begin{table}[h!]
\caption[]{
\label{TABLE01}
    The rmse and maximum error (MaxErr) of the potential energy between  
    the ReaxFF parameters and the QM calculation in this study.
    The labels A--E correspond to the potential energy shown in Figure~\ref{Figure04}:
    the red solid line in Figure~\ref{Figure04} is potential energy 
    using parameter set A. Likewise, green, magenta, sky-blue, 
    and yellow colors are the results using the parameter sets B--E, 
    respectively. 
    The potential energies estimated by the ReaxFF parameters  
    F--H are shown in the SI. 
    rmse and MaxErr are unitless factors because they are normalized 
    by the SD of QM, as shown in eq~\ref{normalizeScore}.
    }
\begin{tabular}{lrr}\hline
    Label &  rmse    &  MaxErr       \\\hline
     A    & 0.100    &  0.577        \\
     B    & 0.107    &  0.840        \\
     C    & 0.114    &  0.495        \\    
     D    & 0.123    &  0.586        \\
     E    & 0.173    &  0.583        \\      
     F    & 0.091    &  0.677        \\
     G    & 0.138    &  0.579        \\    
     H    & 0.104    &  0.536        \\\hline
\end{tabular}
\\
\end{table}%

\begin{table}[h!]
\caption[]{
\label{TABLE02}
    The ratio of six coordination of the Al atom
    during the MD simulation at 2000 K.
    The unit is \%. See the main manuscript for the detailed description.}
\begin{tabular}{lr}\hline
   Label &  ratio of six coordination \\\hline
    A    &    89.69        \\
    B    &    58.92        \\
    C    &    49.33        \\    
    D    &    37.47        \\
    E    &    25.87        \\      
    F    &    48.51        \\
    G    &    62.91        \\    
    H    &    59.57        \\\hline
\end{tabular}
\\
\end{table}%

\begin{table}[h!]
\caption[]{
    The optimized parameters obtained in this study.
    The detail of the physical meaning for ReaxFF parameters such as De(sigma) and p(be$_2$) 
    is shown in ref.~\cite{van2001reaxff}.
    std/avg denotes the (SD/average) \%
\label{TABLE03}}
{\scriptsize
\begin{tabular}{lrrrrrrrrr}\hline
 parameter type  &   A       &   B         &   C         &   D       &   E      &   F      &   G        &   H        & std/avg     \\\hline
 De              & 121.3105  & 102.9408    & 100.976     &  91.03454 & 94.02349 & 95.02255 & 99.82974   &  100.2807  &    8.610    \\
 p(bo$_1$)       & -0.0581   &  -0.0692    &  -0.0578    &  -0.0661  & -0.0705  &  -0.0605 & -0.0598    &   -0.0708  &   -8.237    \\    
 p(bo$_2$)       &  7.9659   &   7.6978    &   8.9451    &   8.2364  &  9.9658  &   8.3418 &  8.6086    &    8.0779  &    7.863    \\
 D$_{ij}$        &  0.0643   &   0.0868    &   0.0678    &   0.0888  &  0.0667  &   0.0937 &  0.0719    &    0.0640  &   15.125    \\
 RvdW            &  1.7782   &   1.7295    &   1.7959    &   1.8925  &  1.8529  &   1.7075 &  1.7065    &    1.7036  &    3.820    \\
 alfa            & 11.1010   &  11.1040    &   11.161    &  10.159   & 10.881   &   11.116 & 11.181     &   11.683   &    3.596    \\
 ro(sigma)       &  1.5421   &   1.5707    &   1.5915    &   1.572   &  1.6638  &   1.5666 &  1.5846    &    1.5703  &    2.117    \\
 p(val$_1$)      & 28.7692   &  33.6726    &  11.134     &  31.3518  & 32.8224  &  34.7920 & 30.8857    &   33.9987  &   24.403    \\
 p(val$_2$)      &  1.4601   &   1.1855    &   1.6583    &   0.8882  &  1.1641  &   1.1993 &  1.4369    &    1.2613  &   17.068    \\\hline
\end{tabular}
}
\\
\end{table}%

\clearpage
\newpage

\begin{center}
  \includegraphics[clip,width=8.0cm]{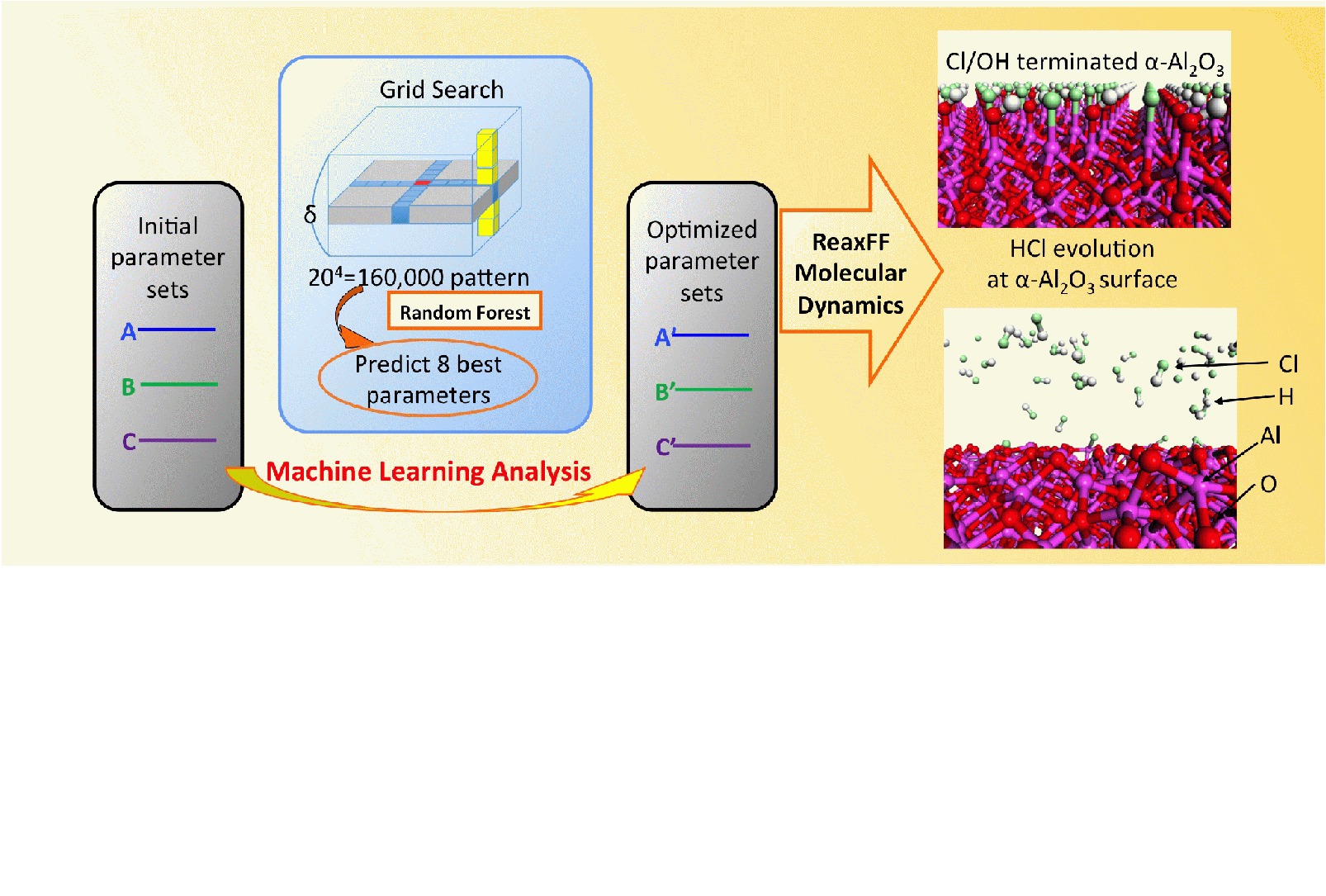} \\
  TOC
\end{center}

\newpage

Figure captions.

\newpage

\newpage

\newpage

\newpage

\newpage

\newpage

\newpage

\end{document}